\documentclass[journal]{IEEEtran}
\usepackage[hyphens]{url}
\usepackage{hyperref}
\hypersetup{colorlinks=false,breaklinks=true}
\usepackage{amsmath,amsfonts}
\usepackage{algorithmic}
\usepackage{array}
\usepackage{xcolor}
\usepackage[caption=false,font=normalsize,labelfont=sf,textfont=sf]{subfig}
\usepackage{textcomp}
\usepackage{stfloats}
\usepackage{url}
\usepackage{verbatim}
\usepackage{graphicx}
\usepackage[utf8]{inputenc}
\usepackage[english]{babel}
\usepackage{amsmath}
\usepackage{array}
\usepackage{amsfonts}
\usepackage{adjustbox}
\usepackage{amssymb}
\usepackage{xcolor}
\usepackage{tabularx}
\usepackage{enumitem}
\usepackage{marginnote}
\usepackage{multirow}
\usepackage{lmodern}
\def\BibTeX{{\rm B\kern-.05em{\sc i\kern-.025em b}\kern-.08em
    T\kern-.1667em\lower.7ex\hbox{E}\kern-.125emX}}
\usepackage{balance}

\usepackage{soul}

\usepackage{booktabs}
\usepackage{footnote}
\usepackage{tabularx}
\usepackage{threeparttable} %
\usepackage{multirow} %
\usepackage{rotating} %
\usepackage{bigstrut} %
\usepackage{pifont}

\newcommand*\OK{\ding{51}}
\usepackage{adjustbox}

\newcolumntype{R}[2]{%
	>{\adjustbox{angle=#1,lap=\width-(#2)}\bgroup}%
	l%
	<{\egroup}%
}

\usepackage{todonotes}
\newboolean{showcomments}
\setboolean{showcomments}{true}         %

\ifthenelse{\boolean{showcomments}}
{\newcommand{\nb}[2]{
		\fbox{\bfseries\sffamily\scriptsize#1}
		{\sf\small$\blacktriangleright$\textit{\textcolor{red}{#2}}$\blacktriangleleft$}
	}
}
{\newcommand{\nb}[2]{}
	
}

\usepackage{scalerel}
\usepackage{tikz}

\usetikzlibrary{svg.path}
\definecolor{orcidlogocol}{HTML}{A6CE39}
\tikzset{
	orcidlogo/.pic={
		\fill[orcidlogocol] svg{M256,128c0,70.7-57.3,128-128,128C57.3,256,0,198.7,0,128C0,57.3,57.3,0,128,0C198.7,0,256,57.3,256,128z};
		\fill[white] svg{M86.3,186.2H70.9V79.1h15.4v48.4V186.2z}
		svg{M108.9,79.1h41.6c39.6,0,57,28.3,57,53.6c0,27.5-21.5,53.6-56.8,53.6h-41.8V79.1z M124.3,172.4h24.5c34.9,0,42.9-26.5,42.9-39.7c0-21.5-13.7-39.7-43.7-39.7h-23.7V172.4z}
		svg{M88.7,56.8c0,5.5-4.5,10.1-10.1,10.1c-5.6,0-10.1-4.6-10.1-10.1c0-5.6,4.5-10.1,10.1-10.1C84.2,46.7,88.7,51.3,88.7,56.8z};
	}
}
\newcommand\orcidicon[1]{\href{https://orcid.org/#1}{\mbox{\scalerel*{
				\begin{tikzpicture}[yscale=-1,transform shape]
					\pic{orcidlogo};
				\end{tikzpicture}
			}{|}}}}

\begin{document}

\title{Railway cyber-security in the era of interconnected systems: a survey}
\author{Simone~Soderi$^{\textsuperscript{\orcidicon{0000-0002-1024-9470}}}$,
Daniele~Masti$^{\textsuperscript{\orcidicon{0000-0002-2889-2309}}}$,
Yuriy~Zacchia~Lun$^{\textsuperscript{\orcidicon{0000-0002-9408-8773}}}$
\thanks{
This work was supported in part by Consorzio Interuniversitario Nazionale per l'Informatica (CINI)  through  Research Project under Grant CA 01/2021 a.i. 2.

Simone Soderi and Daniele Masti are with Scuola IMT Alti Studi Lucca. email: \texttt{\{simone.soderi,daniele.masti\}@imtlucca.it}. 
Yuriy Zacchia Lun is with Università degli Studi dell'Aquila. email:\texttt{yuriy.zacchialun@univaq.it} }}

\maketitle
\vspace{-1.5cm}
\begin{abstract}

Technological advances in the telecommunications industry have brought significant advantages in the management and performance of communication networks. The railway industry is among the ones that have benefited the most. These interconnected systems, however, have a wide area exposed to cyberattacks.

This survey examines the cybersecurity aspects of railway systems by considering the standards, guidelines, frameworks, and technologies used in the industry to assess and mitigate cybersecurity risks, particularly regarding the relationship between safety and security. To do so, we dedicate specific attention to signaling, which fundamental reliance on computer and communication technologies allows us to explore better the multifaceted nature of the security of modern hyperconnected railway systems. 

With this in mind, we then move on to analyzing the approaches and tools that practitioners can use to facilitate the cyber security process. In detail, we present a view on cyber ranges as an enabling technology to model and emulate computer networks and attack-defense scenarios, study vulnerabilities' impact, and finally devise countermeasures. We also discuss several possible use cases strongly connected to the railway industry reality.
\end{abstract}


\noindent\rule{8.4cm}{1pt}
Please cite this version of the paper:\\
S. Soderi, D. Masti and Y. Z. Lun, "Railway Cyber-Security in the Era of Interconnected Systems: A Survey," in \textit{IEEE Transactions on Intelligent Transportation Systems}, doi: 10.1109/TITS.2023.3254442.

You may use the following bibtex entry:
\begin{verbatim}
@ARTICLE{Soderi2023RailwaySurvey,
  author={Soderi, Simone and Masti,
   Daniele and Lun, Yuriy Zacchia},
  journal={IEEE Transactions on 
  Intelligent Transportation Systems}, 
  title={Railway Cyber-Security in the 
  Era of Interconnected Systems: A Survey}, 
  year={2023},
  doi={10.1109/TITS.2023.3254442}}
\end{verbatim}
\noindent\rule{8.4cm}{1pt}

\section{Introduction}

Railways have been one of the main commodities to move passengers and freight since at least the late 19th century. Nevertheless, operators have faced mounting pressure to meet ever-increasing performance and safety demands from the public~\cite{pyrgidis2016railway}. On top of such targets, the awareness of cybersecurity themes has also changed. For these reasons, securing railway systems from cyber attacks has lately become a central issue for practitioners and the public, especially after recent news stories such as~\cite{register2008polish}.

The cause for this abrupt need for answers is simple: while, in the past, railway systems often depended on specifically purposed electromechanical devices that operated in an air-gapped environment, newer infrastructures are often based on commercial-off-the-shelf systems that operate in a fully networked setting. This means that such new installations offer both a much larger attack surface and that attacks can be carried out with shallower knowledge than before. The reliance on shared infrastructures for the operations of multiple subsystems (e.g., both VOIP and signaling might use the same network infrastructure to carry information) amplifies this problem, making the possibility of \textit{lateral movements} extremely relevant. 
This possibility is problematic because railway companies may operate (through the same shared infrastructure) Information and Communication Technology (ICT) services, which have also been hit by various attacks~\cite{Jo2012StudyOnboardICT}. 
This scenario is not unique to the railway sector: many public infrastructures have become victims of cyber attacks in recent years. 

Many proposals have arisen to address this issue. For example, in 2008, the ``European Programme for Critical Infrastructure Protection"~\cite{CouncilEU2008DirectiveCriticalInfra} was established to improve the security of \textit{critical infrastructures}, which are defined as all those systems considered essential to maintaining the vital functions of society.
Recently~\cite{CouncilEU2020DirectiveCriticalInfra}, debates for updating this Directive have restarted to deal with the present threat landscape. The target has been to include a much broader landscape of systems, including the transportation industry and the railway sector, and to champion a novel approach more focused on the resilience of overall integrated infrastructures rather than on the security of individual assets.  

Indeed, considering that a successful attack on railway systems can result in the loss of the safety guarantees of the network~\cite{Milligan2014SecuringRailwayControl}, the rail transportation sector cannot ignore cybersecurity anymore and must start considering cybersecurity, physical security, and safety as intertwined aspects that cannot be dealt with separately. 
For these reasons, developing a new generation of methods for verifying and hardening rail systems has become of great practical and theoretical importance.

In this survey, we investigate how the industry has responded to such a challenge by:

\begin{itemize}
\item {collecting the standards governing the many safety-critical subsystems that make up a complete railway network;}
\item {recalling some of the most significant cyberattacks carried out in recent years on railways systems;}
\item {investigating the cybersecurity projects involving railway signaling systems}
\item {investigating an approach based on \textit{cyber ranges} to emulate and verify the security of networking systems similar to those used in the railway industry. }
\end{itemize} 

The paper is organized as follows: in Section~\ref{sec:integratedSys}, we recall the main components of a railway system; in Section~\ref{sec:facets}, we introduce the facet of security in the railway sector in general and, in Section~\ref{sec:signaling}, we discuss the relationship between safety and security. %
Section~\ref{sec:risk-mgmt} concludes with a novel methodology for performing cybersecurity assessments.
In Section~\ref{sec:ranges}, we report how cyber ranges can be valuable in performing cybersecurity assessments and propose some railway-centered scenarios that might be of interest for discussion and future work. In Section~\ref{SEC:Conclusions}, we finally draw some final remarks and discuss further developments.

\section{Railways as an integrated system}
\label{sec:integratedSys}

In a broad sense, railways can be defined as a collection of different systems whose purpose is to transfer passengers and goods on wheeled vehicles running on rails located on tracks. Such subsystems can be broadly collected into three families: 
\begin{itemize}

\item \textbf{Railway infrastructure} comprises all the tracks (sometimes referred to as the \textit{permanent way}), all the civil works, and the systems and premises that ensure the regular traffic flow. In literature, this latter component is often further divided to distinguish between the so-called ``facilities and premises,'' which encompass stations, depots, and other similar facilities and  \textit{wayside systems} that operate along the lines, which encompass signaling systems, electrification facilities (which hardening is deeply interlaced with the security of the electrical grid as a whole~\cite{gazafrudi2014power}) and level crossings.

\item \textbf{Rolling stocks} comprise powered vehicles (locomotives, single rail cars, shunters, etc.), engineering vehicles, and trailer vehicles.

\item \textbf{Railway operations} encompass the technical duties performed to ensure trains circulate and the commercial operations that railway companies perform to ensure revenue ~\cite{pyrgidis2016railway}. 

\end{itemize}
Most tasks carried out in a railway company involve all those three subsystems simultaneously. This suggests that ``holistic'' approaches that favor securing the system as a whole~\cite{thaduri2019cybersecurity} should be preferred to approaches that focus on securing a single component of the system without caring for its overall capability to accomplish its many tasks.

\subsection{The dualism between safety and security} 
\label{subsec:safety-security}

Safety is deeply ingrained into modern industry, and railway makes no exception. Indeed, railways and other transportation systems are classified as safety-critical since their failure may result in loss of human life or disasters of another sort. The design of this kind of system has traditionally followed a ``safety above all'' paradigm, meaning that, to be considered fit to be used, each component (and the system as a whole) must achieve a minimum Safety Integrity Level (SIL)~\cite{cenelec2018en50126,iec2010Standard61508}. This means that specific design rules and test procedures must be implemented following a specific set of standards and norms, guaranteeing that the system continues to fulfill its safety requirements even in case of random failure. 

Nevertheless, despite the observation that an insecure system has much fewer chances to behave in a safe manner, many widely adopted safety standards do not consider cybersecurity explicitly or, at most, only generically mention that implementers should include cybersecurity mechanisms~\cite{Hansen2009SecurityAttackAnalysis, Gronbaek2008SafeWireless, valdivia2018cybersecurity} in their design\footnote{We will come back to this issue later in the paper.}.

This lack of guidance in such an otherwise pervasive recommendations framework, coupled with the high cost and the relative inapplicability of otherwise commonly adopted security frameworks in railway scenarios, however, has often pushed companies to consider security as an after-thought of the overall design process of new railways systems, an after-thought often delayed due to business and cost reasons~\cite{wang2022CPScybersecurity, habibzadeh2019survey}. Such a ``lazy'' approach is perhaps even more surprising considering that the relative simplicity of accessing the infrastructure~\cite{kiviharju2022cryptographic}, coupled with the vast effect that successful attacks on railway infrastructures may cause on the public at large, makes the railway infrastructure a juicy target for all kinds of attackers, from state-sponsored actors down to ``script kiddies.''
This strategy may even end up causing more problems in the long term since security specialists may have to work on infrastructures composed of a complicated landscape of systems whose overall functioning is linked to insecure-by-design architectures, possibly too old to be coupled with modern security solutions.

Indeed, railway infrastructure has been the subject of numerous attacks in recent years. In Table~\ref{table:incidents}, we summarize a few significant confirmed cybersecurity incidents that have affected or had the potential to compromise transportation operation safety. Looking at the Table,  it is easy to see that while the operations and safety systems were the primary targets in the earliest incidents, the attackers’ focus has shifted mainly toward ITC-related systems in recent times. 
Nonetheless, many recent attacks still significantly disturbed the transportation services as a whole, possibly due to undisclosed (or possibly even the simple fear of) lateral movement by the attackers.

\subsection{Security landscape in the railway industry}
\begin{table*}[t!]%
	\centering
	\begin{threeparttable}
		\renewcommand{\arraystretch}{1.2}
		\caption{Timeline of cybersecurity incidents in the railway sector with their description.}\label{table:incidents} 
		\begin{footnotesize}

        \begin{tabularx}{\linewidth}{| r| X |}
         \hline
        ~~\textbf{Date} & \textbf{Description} \\
         \hline
		{August 2003} & A computer virus disabled the CSX Transportation headquarters in Florida, affecting signaling in thousands of km of railway line. This incident has also been referred to as the ``Sobig'' incident~\cite{cbs2003virus,temple2016railwayfailure}.\\

		 \hline
        {January 2008} & A teenager derailed four tram vehicles causing injuries to twelve people after hacking a train network of Lodz, Poland~\cite{register2008polish}.\\

         \hline
        {December 2012} & A cyberattack on a Northwestern US rail company's computers disrupted railway signals for two days~\cite{wired2012hackers}.\\
        
         \hline
        {March  2015} & The HoneyTrain Project recorded over two millions logins attempts with four successful illegal accesses to the Human-Machine Interface (HMI) of a virtual train control system in the space of six weeks~\cite{honeytrain2015,heise2015honeytrain}.\\
        
         \hline
        
        {Febraury  2016} & BlackEnergy and KillDisk malware infected the systems of a prominent Ukrainian rail company. In December 2015 the Ukraine power grid cyberattack was also attaked using the same malwares~\cite{trendmicro2016killdisk}.\\
         \hline
         
        {July  2016} & A study reported that the UK Network Rail had been hit by at least four significant cyberattacks over 12 months, including intrusion in rail infrastructure itself. According to such a study, these attacks seemed to be exploratory~\cite{sky2016four}. \\
        
         \hline
        {November  2016} & A ransomware attack took ticket machines of the San Francisco light rail transit system (SF Muni) offline for a day,  There was no impact on transit service, the safety systems, or customers' personal information~\cite{wsj2017germany,guardian2016ransomware}. \\ 
        
        \hline     
        {May  2017} & Deutsche Bahn, suffered a ransomware attack on its data systems~\cite{thelocal2017international}. The same computer virus also hit the national railway systems in Russia~\cite{tass2017virus} and China~\cite{apnews2017latest}.\\
        
         \hline
         
        {October  2017} & Sweden's transportation Administration was targeted by a DDoS attack on the IT systems that monitor railway traffic. Two DDoS attacks hit the public transportation operator V\"{a}sttrafik the next day~\cite{local2017swedish}.\\
        
         \hline
        {May  2018} & The Danish operator DSB came under a DDoS attack, making it impossible to purchase tickets. Internal mail and telephone systems used by the DSB staff were also affected~\cite{local2018danish}. \\
        
         \hline        
         
        {March 2019} & An Israeli cyber threat intelligence company identified an actor operating on a top-tier dark web forum selling access to an administrative panel of a Chinese rail control system~\cite{sixgill2019chinese}. \\
        
         \hline
        {October 2020} & A ransomware attack hit the Soci\'{e}t\'{e} de transport de Montr\'{e}al (STM) compromising 624 operationally sensitive servers. The outage also affected STM for over a week~\cite{cbc2020stm,bc2020montreal}.\\
        
         \hline
         
        {December 2020} & A ransomware attack hit OmniTRAX. It was the first publicly disclosed case of a so-called double-exhortation ransomware attack against a US freight rail operator~\cite{fw2020ransomware}. \\
        
         \hline
        {December 2020} & The Egregor ransomware attack hit TransLink, forcing the company to shut down several IT services including part of payment systems~\cite{zdnet2020ransomware}. No transit safety systems were affected, but the IT problems impacted GPS functions on buses~\cite{news2020translink} and information regarding personal banking social insurance information may have been compromised~\cite{gn2020translink}\\       
        
        \hline
        {July 2021} & A cyberattack on Iran's railroad system caused chaos across the whole country~\cite{reuters2021iran}.\\
        
         \hline
        {October 2021} & The Toronto Transit Commission (TTC) became a victim of a ransomware attack, losing access to systems used to communicate with vehicle operators, online booking, etc.~\cite{gn2021ttc}. Subsequently, the TTC announced that personal information of (former) employees, may have been stolen~\cite{gn2021update}. \\
        
         \hline
        {March 2022} & Italian Railway Operator Trenitalia and National infrastructure holder were affected by a ``cryptolocker infection'', causing disruption of service~\cite{reuters2022Trenitalia}. \\
        
                 \hline
        {April 2022} & Linked to events in the Russian-Ukranian conflict, ``\textit{a clandestine network of railway workers, hackers and dissident security forces went into action to disable or disrupt the railway links connecting Russia to Ukraine through Belarus}''~\cite{WP2022Bielorussia}. \\

         \hline         
        \end{tabularx}
		\end{footnotesize}
	\end{threeparttable}
\end{table*}

The importance of the target and the relative lack of existing approaches to railway security has pushed many authors to propose analysis for various kinds of attacks and their possible mitigations, also within the academic community. We refer to Appendix~\ref{appendix:lista} for a brief list of methodologies that can be used to analyze cybernetics attacks and to~\cite{wang2022CPScybersecurity,kour2022review} for an extensive analysis of many works centered around railway-specific scenarios.

Nevertheless, as Wang \textit{et al.}~\cite{wang2022CPScybersecurity} and Kour \textit{et al.}~\cite{kour2022review} also report, the analyses done to this day are too often concerned with particular aspects of a single system. In other words, they work without caring for the overall picture and context in which they are inserted, thus not considering how interactions between coupled systems may amplify or negate some threats. In this era where railway systems are made of tens of separate interconnected safety and security-critical systems, this approach may result in analysis and countermeasures of limited applicability and effect.

This problem cumulates with the already mentioned lack of cybersecurity awareness in the often legally-binding standards used in the railway sector. To give some examples, the CENELEC EN 50159~\cite{cenelec-en-50159}, which is used to design communication between safety-related equipment, addresses topics such as message authenticity and integrity. However, it does not cover general cybersecurity issues like preventing overloading transmission systems or ensuring the confidentiality of safety-related information. 
Another example is the IEC 61508~\cite{iec2010Standard61508}, which can be considered the general standard for achieving the safety of electronic devices and is extensively used in the railway industry, which does not cover security issues~\cite{valdivia2018cybersecurity}. Indeed both standards only mention that intentional malicious human actions must be considered and generically refer to the ISA/IEC~62443~\cite{isaiec2009Standard62443} standard.
A similar landscape can also be found concerning the technical norms governing control platforms doors and wayside control systems. For the former, the primary reference is the GB~50157-2003~\cite{gbNational2003Standard50157}, which again does not tackle security issues~\cite{valdivia2018cybersecurity}.

\begin{figure}[t]
	\centering
	\includegraphics[width=0.85\columnwidth]{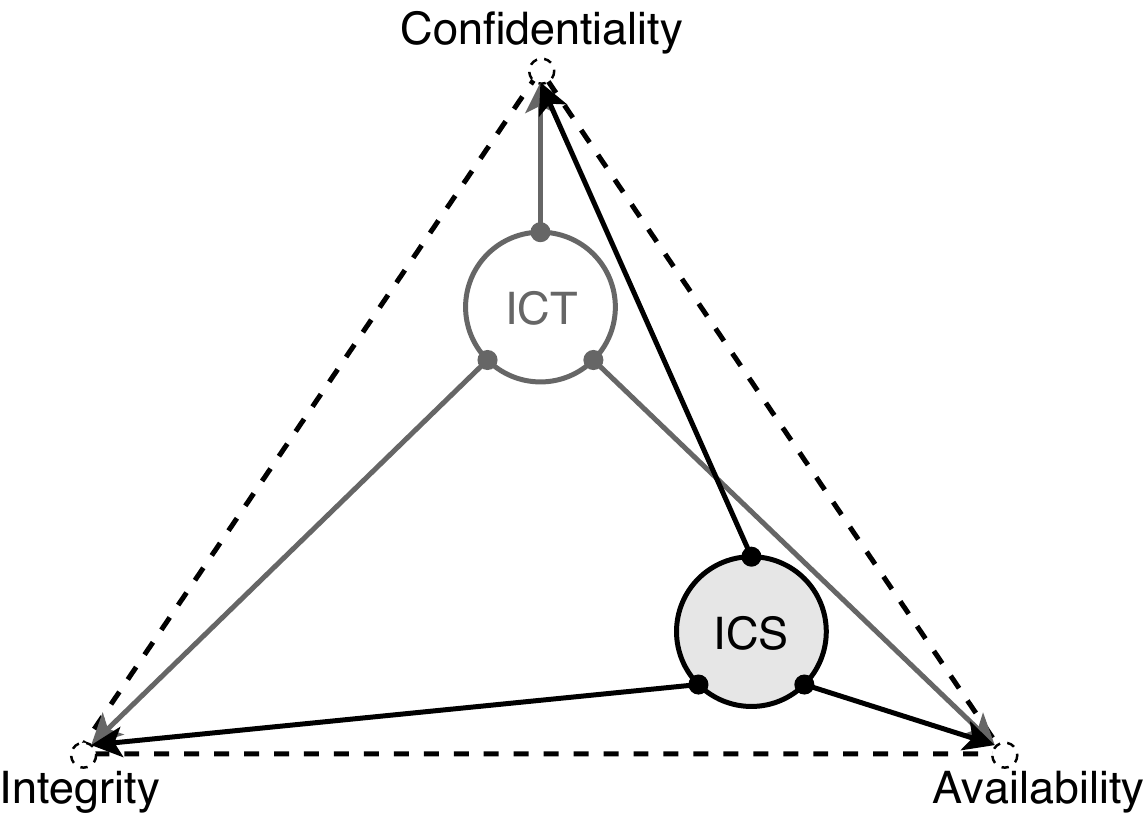}
	\caption{The different meanings of the Confidentiality, Integrity, and Availability triad in ICT and IC systems.}

	\label{fig:cia}
\end{figure}

Authors and governing bodies tried to address this situation, yet before analyzing their proposal, it is meaningful to analyze the unique requirements of railway systems compared to ICT systems.
Nowadays, railway projects heavily rely on classical Industrial Control Systems (ICS) to control electromechanical systems and automate industrial processes and operations in various applications. Such systems often include programmable logic controllers, data communication systems, and supervisory control and data acquisition. Securing IC systems poses different challenges than securing pure ICT systems. Consider, for instance, the Confidentiality, Integrity, and Availability triad, a well-known model that defines the security requirements to support organizations in specifying the core security objectives of their systems~\cite{2008:securityBook}. As shown in Figure~\ref{fig:cia}, while ICT security focuses on confidentiality to prevent stealing private information, ICSs are more concerned with data integrity and avoiding unplanned system outages that can disrupt production availability and profitability.

Communication between subsystems also plays a central role and is usually achieved through a transportation network managed via a central Operation Control Center (OCC), where many operational tasks are merged.
Currently, there is no consensus about how to design such control centers~\cite{valdivia2018cybersecurity}, and many different OCC configurations have been designed following possibly incompatible standards. Among them, the APTA RT-OP-S-005-03~\cite{apta1018StandardRT} is among the most used ones, yet it considers only physical security and provides no guidelines for cybersecurity. Moreover, compared to classical ICSs, the transportation sector poses even more importance to the concept of resilience~\cite{levy2015cyber} since the availability of each subsystem has a paramount priority.

There is also the question about \textit{how} connectivity should be offered to each one of such components: in theory, railway communications could build on many different technological options. Nevertheless, while the use of dedicated technology was practically mandatory in the past, in recent times, there has been a solid push for the use of non-dedicated backbones and off-the-shelf technologies to reduce setup and recurring costs and speed up deployment~\cite{kiviharju2022cryptographic}. Indeed, although modern connectivity technologies like the 5G offers many security features, it is yet to be understood if (and how) they can be adapted to the railway's needs.
In the next Section, we will analyze some of these aspects in more detail.

\begin{figure*}[t]
	\centering
	\includegraphics[width=0.98\textwidth]{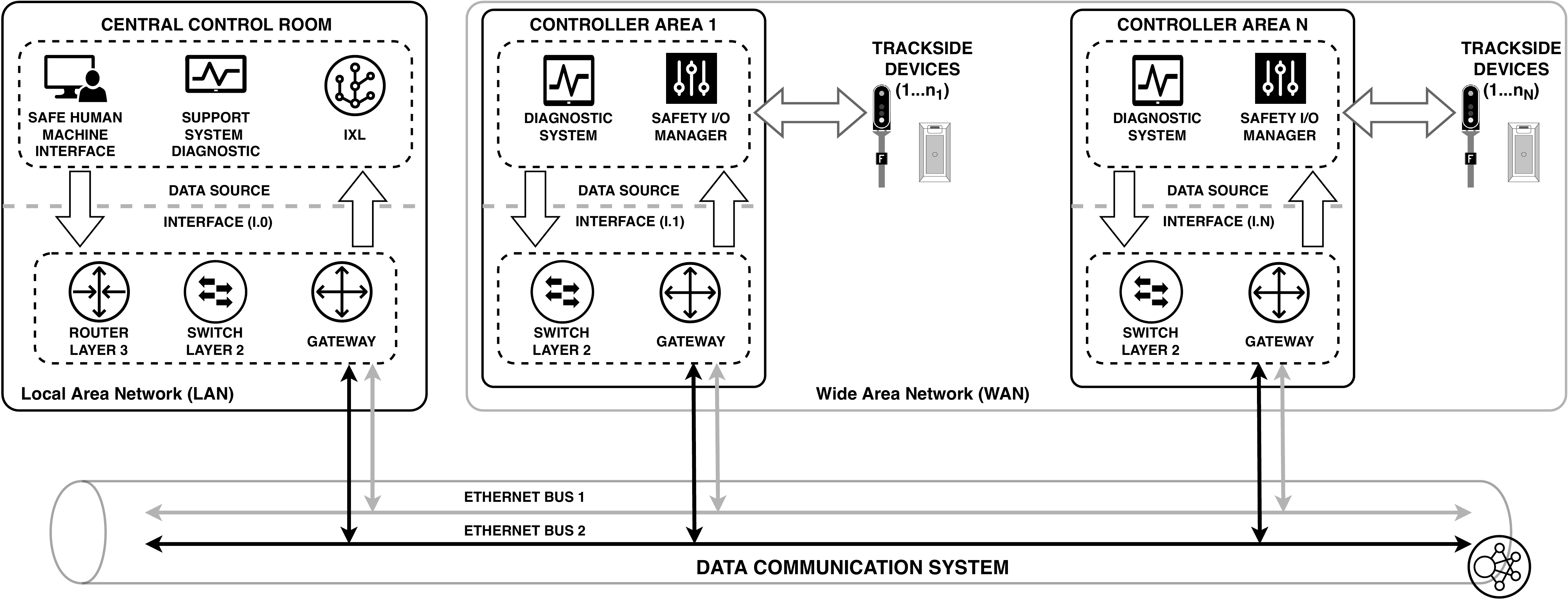}
	\caption{Wayside network scenario. The central control room connects the safe HMI with support system diagnostic and interlocking in LAN and controller areas via WAN. Each LAN comprises networking devices such as gateways, routers, and switches. The control areas link trackside devices to their safety input/output manager and diagnostic system.}
	\label{fig:wayside-scenario}
\end{figure*}

Figure~\ref{fig:wayside-scenario} depicts the interconnected nature of modern railway systems. In there, we can see how the functioning of railway subsystems is assured by a vast number of devices positioned along the line, all reporting to operators and central systems located in central control rooms. The same picture also shows how the connectivity between all these devices is potentially offered by means of a network infrastructure whose fundamental principles are not necessarily much different from the ones used in classical enterprise ICT settings. Such a network backbone is also possibly shared with other railway or ICT subsystems. In such a case, logical separations between the data fluxes can be assured by the use of Virtual Private Network (VPN) technology.

\subsection{A deeper look on signaling systems}

One of the systems where the role of connectivity is the most pervasive is signaling. For this reason, we will take this primary function, and European Rail Traffic Management System (ERTMS) in particular, as the case study for the rest of the paper.

Signaling comprises all the machinery necessary to ensure the safe movements of rolling stocks on railway infrastructure~\cite{pachl2020railway} and is part of the so-called wayside systems. 
In detail, signaling systems are comprised of a few main components tasked to:
\begin{itemize}
\item check the clearance of track sections using either track circuits or axle counters;
\item lock movable track elements such as switches and crossings in a proper position;
\item prevent conflicting train movements through the action of an interlocking system. This system is responsible for granting a train exclusive access to a \emph{route}, which is a sequence of track elements exclusively assigned for train movement through a station or a junction~\cite{fantechi2019ConnectedTrains};
\item controlling railway vehicles to keep them safely apart and within speed limits through Automatic Train Control (ATC) systems.
\end{itemize}  
ATC systems can be further divided into three subsystems: Automatic Train Protection (ATP), Automatic Train Supervision (ATS), and Automatic Train Operation (ATO)~\cite{ieee2005Stadard14741}. ATP is a vital subsystem that continuously ensures compliance with the maximum safe speed and minimum safe distance limits. ATS often acts upon the signals generated by the interlocking system to monitor and adjust the performance of individual trains to ensure smooth railway service.
The ATO subsystem performs those functions otherwise assigned to the train operator and meets all operating conditions and limits set by the ATC, following the requirements of the railway system to ensure passenger comfort by establishing policies for safe operations ~\cite{fantechi2019ConnectedTrains}. 
All modern railway signaling systems, such as the European Train Control System (ETCS) and Communications-based train control, include ATP functions~\cite{ieee2005Stadard14741}.

Speaking about ETCS, such a system is used as the signaling and control component of the ERTMS, which is the \textit{de facto} global standard~\cite{ghazel2017control} in the high speed and mainline railway market segment (please refer to Appendix~\ref{sec:market} for an overview of the railway market). 

The ERTMS standard has been designed to be an almost universal traffic management solution and specify several service levels, each of which enables more and more tasks to be accomplished by the system. The functions offered by different ERTMS range from none (in the case of a ``Level 0'') to a complete virtual train coupling system~\cite{schenker2020s2r}, which is the main subject of the possible future  ERTMS/ETCS ``Level 4''.

This potential, however,  has a cost in terms of the complexity and required capabilities of the employed communication channel. This, in turn, determines the type of equipment to be used~\cite{pachl2020railway}.
At Level 1, for instance, the system relies on \textit{intermittent} ATP architecture that uses controlled transponders (``balises''\footnote{Eurobalise Transmission System is a safe spot transmission-based system conveying safety-related information between the wayside infrastructure and the train~\cite{UNISIG2012FFFIS}} or loops) positioned along the tracks. Such devices relay information received via a traditional signaling system via a Line-side Electronic Unit (LEU). At Levels 2 and 3, instead, the ETCS works as a \textit{continuous} ATP system, which requires bidirectional Vehicle-to-Infrastructure (V2I) communication. In this case, railway cabs receive information from balises, short loop antennas, or digital radio\footnote{ Interestingly, thanks to the strong push on commonality, all ETCS levels use the same onboard equipment.}.

To this date, the technology used as digital radio is the GSM-R system, which has been built on top of 2G GSM cellular technology.
However, the GSM-R technology is starting to show its limits as it cannot provide enough support for growing ERTMS demands for autonomous train operation capabilities. For this reason, there has been a push to move to more advanced (possibly packed-switched) technologies.
In this sense, a natural candidate would have been the LTE-R~\cite{mikulski2018management} (based on 4G cellular infrastructure), which has found successful applications in the Asian markets. However, as trains move faster and the quantity of data to transmit grows, it is easy to imagine scenarios in which even the current 4G technology would fall short. 
The novel ``Future Railway Mobile Communication System'' (FRMCS)~\cite{uic2020frmcs}, which relies on 5G technology, will probably be able to solve these issues. Nevertheless, in the meantime, the already mentioned LTE-R or the Finnish national ``terrestrial trunked radio standard''  (TETRA)~\cite{ETSI2011TETRA,moreno2015survey} are being adopted, even if only as stop-gap solutions.

\subsection{Security aspects of ERTMS}

The central role of ERTMS (and signaling systems in a broad sense) in guaranteeing a safe circulation of trains has sparked the academic community to assess its security properties.

For instance, in~\cite{chothia2017attack}, the Authors highlight weaknesses of GSM-R and EURORADIO protocols that would allow an attacker to forge train control messages. Although the Authors recognize that it would be challenging to carry out this kind of attack in an environment where only small segments of data are exchanged, their hypothetical attack will become a more pressing possibility in a context where the quantity of exchanged data will grow.

Another example is the attack proposed in~\cite{bloomfield2012secure}, in which the authors exploit the fail-safe behavior of ERTMS to engineer a Denial-of-Service attack that causes a train to halt. Although the authors also argue that causing an accident might be possible, as shown in~\cite{wang2022CPScybersecurity}, also causing a delay for a single train can be enough to cause severe repercussions on the overall railway schedule and service.

More theoretical approaches have also been used to test the protocol's security (and safety). In~\cite{ruiter2016formal}, a formal analysis of the train-to-trackside communication protocols used in the EURORADIO protocol is carried out using pi-calculus and the ProVerif tool. In there, the Authors found out that the protocol is not secure against forging of emergency messages in all those cases in which the underlying carrier network is compromised (such as by an IMSI catching in the case of GSM-R), and it is also weak against  message deletion, and downgrading attacks.

Many authors have also proposed possible mitigation of known problems of these protocols. For instance, the very Authors of~\cite{ruiter2016formal} make several recommendations further to enhance the security of ERTMS in the same paper. In~\cite{chen2011performanceTCS}, another approach to improve the protocol in EURORADIO is proposed by the Authors, accompanied by theoretical consideration employing Colored Petri Networks to verify the superior performance (both in safety and security terms) of their proposal.

Cryptographic considerations on the EURORADIO protocol and its coupling with various digital radio technologies have also been performed. In~\cite{kiviharju2022cryptographic}, for instance, it has been argued that the 3DES-based approach of EURORADIO is not secure enough given the nowadays available computational power. Attacks of this kind, however, often rely heavily on intrinsic properties of the GSM-R carrier and are thus heavily countered by the use of more modern carriers. We refer the interested reader to~\cite{kiviharju2022cryptographic} for more in-depth considerations regarding the improved security primitives offered by LTE-R, 5G, and FRMC.

\begin{figure*}[t]
	\centering
	\includegraphics[width=.98\textwidth]{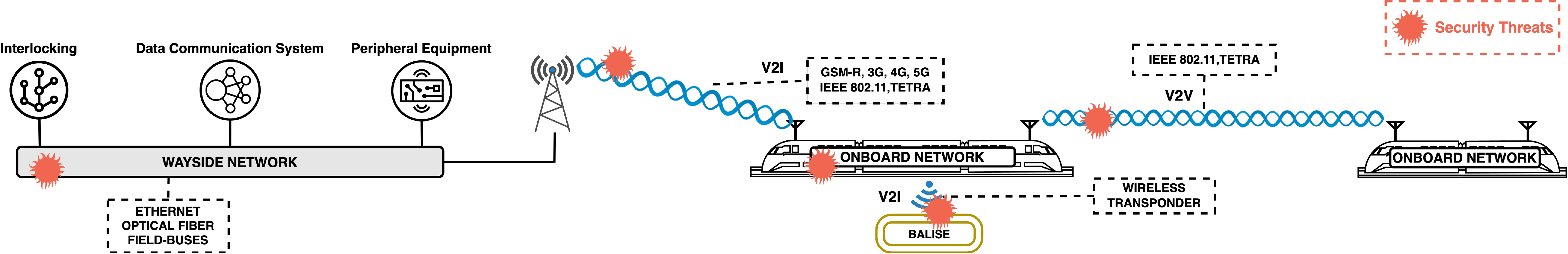}
	\caption{
	A schematic representation of the many communication channel used by railways systems. %
	Dashed boxes list possible technological solutions.  Each one of those links can be be potentially used to to carry out attacks to the connected subsystems.%
	}

	\label{fig:signaling-scenario}
\end{figure*}

\subsection{The limits of an atomistic approach}

As already noted, all these analyses rarely take into account the overall placement of the analyzed system into the ``operational pipeline'' of a railway system. For this reason, they fall short of identifying the effects of a given attack on the railway system as a whole. Although acceptable in an academic setting, this approach is very myopic from a risk management point of view. 

In Figure~\ref{fig:signaling-scenario}, we show a schematic representation of a railway system. Each component and each V2I, vehicle-to-vehicle, and spot communication channel can be subject to security threats. Wireless communication offers new possibilities for support and new services and also increases complexity during development as it exposes a broader attack surface.
This picture is helpful to see how securing a single subsystem without regard for its placement in a general scheme may not achieve desirable overall security characteristics.

In an enterprise, there is also the urge to \textit{quantitatively} investigate the effect of a given event. In systems as complex as railway signaling infrastructure, however, this feat is well beyond the possibility of a pencil-and-paper-based approach. Indeed, as noted in~\cite{wang2022CPScybersecurity}, simulators and verifiers have become invaluable tools for such a feat. Later in the paper, we will explore how \textit{cyber-ranges} can be used to investigate how network conditions can propagate through a communication backbone and possibly disrupt the services which rely on it to accomplish their function.

\section{Security assessment methodologies for railway systems}
\label{sec:facets}

Although we cited only a few examples of possible attacks on a particular subsystem and cited just a few news events regarding successful attacks, it should be clear by now that railway players cannot simply ignore the security facets of the systems they operate. 

As also mentioned, compared to a purely academic setting, the concept of \textit{governance} (i.e., procedures, management, and certifications) assumes a more prominent role in an enterprise setting compared to the sheer discovery of novel vulnerabilities. This is even more true for companies operating in a setting where safety management has always played a primary role.

Given this premise, it should be no surprise that in recent times there has been a strong push also for developing cybersecurity \textit{risk management} procedures. 

In the remaining of this Section, we report some of the most relevant industry standards that one can apply in the rail industry and then introduce the general ideas behind these industrial schemes\footnote{Nevertheless, it must be noted that recently also Academia has started answering to this call as well. See, for instance, \cite{wang2022CPScybersecurity}.}. Later we will also introduce an original approach to achieve such a feat.

\subsection{ Standards for cybersecurity assessments}
\label{subsec:cybersec}

Currently, procedures for security assessment of railway systems are mostly framed within the ISA/IEC 62443 standard~\cite{isaiec2009Standard62443}, which is the global standard for network security of industrial control systems. Such a document guides ICS  operators through a pipeline that establishes all the requirements, controls, and best practices necessary for securing industrial networks.

Other generally applicable norms and frameworks regarding cybersecurity are:
\begin{itemize}
\item  the Common Criteria for Information Technology Security Evaluation, also known as ISO/IEC 15408~\cite{iec2009Standard15408}. This standard introduces security specifications, implementation, and evaluation procedures tailored for the designated use environment;

\item the ISO/IEC 27001~\cite{isoiec2013Standard27001} standard, which specifies requirements for establishing, implementing, and maintaining information security management systems;

\item  the Cybersecurity Framework (CSF)~\cite{Barrett2018NISTframework}. CSF consists of standards, guidelines, and best practices related to cybersecurity risk management. It also provides a common language for communicating cybersecurity expectations and awareness within and across organizations.

\item the Open Source Security Testing Methodology Manual~\cite{Herzog20120OSSTMM} (OSSTMM), which is the  de facto standard for vulnerability assessment thanks to an auditing methodology aimed to satisfy regulatory and industry requirements.

\end{itemize}

All those standards and frameworks, however, are rather generic and not tailored to the needs of railway systems. To address this issue, the European Networks and Information Systems Agency has established a series of specific security requirements and measures for the operators of essential services that can be recast in the frameworks mentioned above~\cite{ENISA2017Mapping}. %

\subsection{Guidelines to enhance the security of railway systems}

Players like the UK Department for Transport, the International Union of Railways, etc., have also produced some guidelines especially centered on the security facets of railway systems~\cite{dft2016guidance,uic2018guidelines,antoni2018argus}. 
Bloomfield~\textit{et al.}~\cite{bloomfield2016risk} also provided a high-level cybersecurity risk assessment procedure for generic ERTMS-based railway infrastructures and ETCS onboard systems.
Several projects have also tried to address the rail sector cybersecurity challenges under the Shift2Rail~\cite{masson2017cyber} initiative, a European public-private joint undertaking for rail research.
In particular, the two arguably most significant projects under this umbrella have been:
\begin{itemize}
\item  4SECURail~\cite{4SECURail}, a project that addresses the call for formal methods in the railway environment and supports implementing a computer security incident response team for railways;
\item CYRail, which has produced various guidelines to enhance the security of railway systems~\cite{cyrail2017SafetySecurityRail};
  \end{itemize}
but also X2Rail-1~\cite{X2rail}, Roll2Rail~\cite{Roll2Rail}, and Safe4RAIL~\cite{safe4rail,rekik2018cyber} projects are worth to be mentioned.

Moreover, we cannot not mention the NIST Special Publications Series 800, particularly the NIST SP 800-53~\cite{NIST2020Series80053}, which includes the NIST CSF security controls. We also cite the NIST SP 800-82~\cite{Stouffer2015NIST80082} deals with ICSs security controls often used for security in railways.

\subsection{The issue of safety certification}
\label{sec:security_safety}

Given the complicated and hyper-connected nature of modern railways systems, it is no surprise that companies have adopted a landscape of solutions relying on standards both from ITC and Operation Technology (OT) domains to secure their systems. Although this approach follows a trend already in use in other sectors, such as avionics and automotive, it also poses clear challenges in integrating all the prescribed guidelines in the same design. 

In Table~\ref{tab:standards}, we summarize some of the principal design-oriented guidelines currently in use in the railway industry. There, as anticipated, we can see a landscape of safety and security standards, which in some cases also have to coexist at a very intimate level. This need arises, for example, in the devices involved in signaling subsystems. In there, for instance, the object controller of a trackside device (see Figure~\ref{fig:wayside-scenario}) will necessarily have to work both in a fail-safe but also very secure manner.

Moreover, the issue of safety certification still stands: since no standard guidelines to certify the safety of security modules exist, certifying and including security hardware/software in actual railway projects is far from trivial. 
To address this problem, some Authors~\cite{valdivia2018cybersecurity} suggested that manufacturers should physically separate the security modules from the safety modules.%
The novel CENELEC TS 50701 ``Railway Applications – Cybersecurity''~\cite{cenelec2021ts50701} is possibly the first attempt from a standardization body to solve such an integration issue.
This technical specification is based on ISA/IEC 62443 and provides a tailored solution for the railway industry, including rolling stock, signaling, and infrastructure. CENELEC will assess this document in three years and possibly transform it into a standard~\cite{Schlehuber2021cenelec}.

Given the complexity of the safety process defined by EN~50126~\cite{cenelec2018en50126} and of the cybersecurity process described by TS 50701, however, it is imaginable that the synchronization between safety and security will be pretty complicated, especially considering that the two processes certainly have different duration and that cybersecurity management is also a practically never-ending process. In addition, the system under consideration defined by TS~50701 might have a perimeter of application concerning the safety process.

These facts call for a radical shift into the working pipeline usually adopted by the railway industry as they  make it basically mandatory  to consider the security requirements of the final product from the very beginning of the design process.%

\begin{figure}[!t]
	\centering
	\includegraphics[width=0.99\columnwidth]{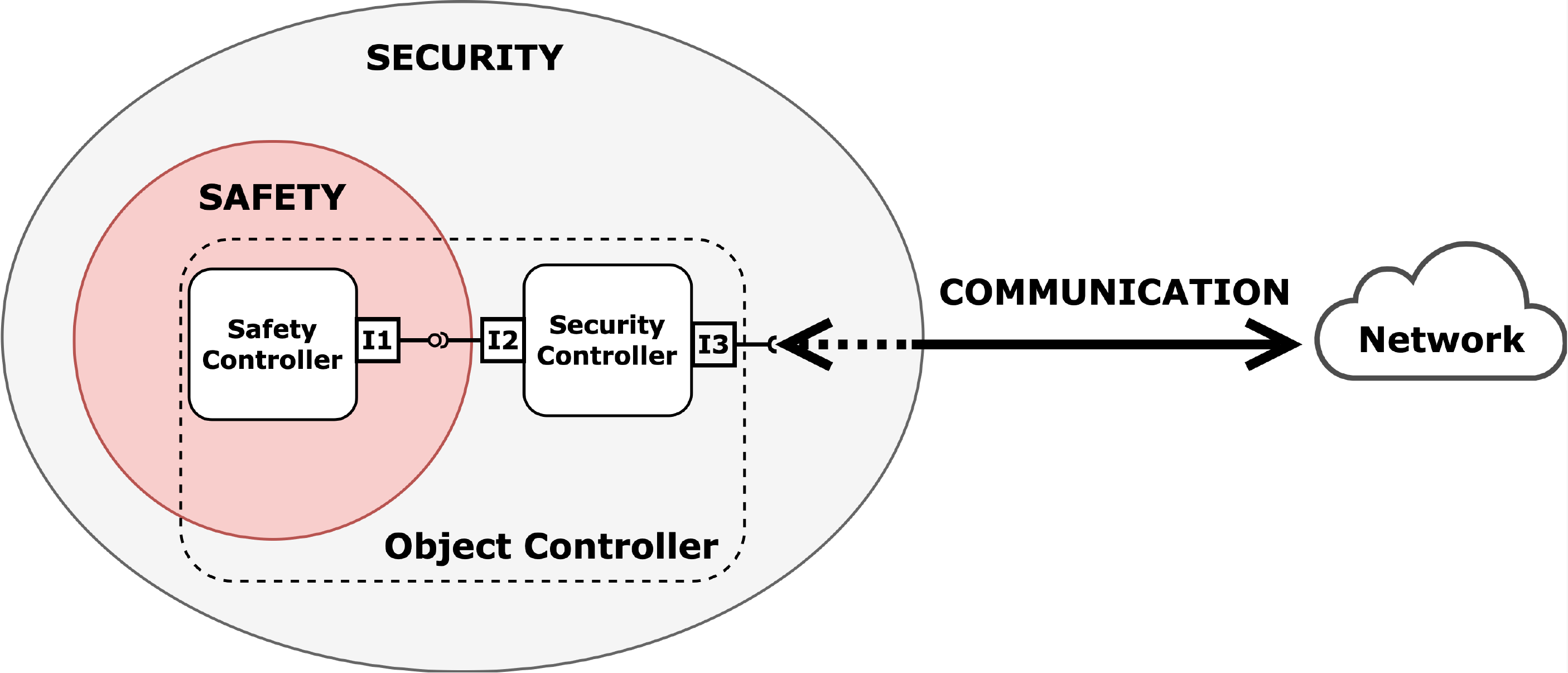}
	\caption{ Example of a possible design architecture integrating both Safety and Security facets. In this approach, the security facets functions as external shell protecting safety function.}
	\label{fig:safety-security-shell}
\end{figure}

A possible architecture that may be used to meet these requirements is shown in Figure~\ref{fig:safety-security-shell}. In such a design, a security \textit{shell} protects the safety function~\cite{2017:security-shell}. This leaves the designer free to apply any relevant standard (possibly from Table~\ref{tab:standards}) to design each internal component but imposes that all communications must go through a security controller, which will also function as the only interface to its safety counterpart. Such an architecture is implicitly resistant, for instance, to a DoS attack because, even in the worst case, only the functioning of the security controller can be compromised, thus leaving any internal fail-safe mechanism intact. 
In addition, this architecture allows one to combine safety and security in a very streamlined way since the designers only have to worry about maintaining the (reciprocal) compatibility between the $I1$ and $I2$ interfaces that connect the two controllers.

\begin{table}[t]
\centering
\caption{Remarks on the applicability of standards, frameworks and guidelines in the railway industry.}
\label{tab:standards}
\begin{tabular}{@{}cl*{3}c}
& \begin{tabular}[l]{@{}l@{}} \textbf{Standards } \\ \textbf{and  guidelines }\end{tabular} & \begin{tabular}[l]{@{}c@{}} \textbf{Application} \\ \textbf{area} \end{tabular} & \textbf{Security} & \textbf{Safety}  \\
\cmidrule{2-5}
& \begin{tabular}[l]{@{}l@{}} ISO/IEC 15408~\cite{iec2009Standard15408}\end{tabular} & IT & \OK &    \\ \cmidrule[1pt]{2-5}
& \begin{tabular}[l]{@{}l@{}} ISO/IEC 27001~\cite{isoiec2013Standard27001}\end{tabular} & IT & \OK  &     \\ \cmidrule[1pt]{2-5}
& \begin{tabular}[l]{@{}l@{}} ISO/IEC 62443~\cite{isaiec2009Standard62443}\end{tabular} & OT &\OK   &    \\ \cmidrule[1pt]{2-5}
& \begin{tabular}[l]{@{}l@{}} CLC/TS 50701~\cite{cenelec2021ts50701} \end{tabular} & OT & \OK  & \OK     \\ \cmidrule[1pt]{2-5}
& \begin{tabular}[l]{@{}l@{}} NIST CSF~\cite{Barrett2018NISTframework} \end{tabular}     & IT, OT &\OK   &  \\ \cmidrule[1pt]{2-5}
& \begin{tabular}[l]{@{}l@{}} NIST 800-82~\cite{Stouffer2015NIST80082}\end{tabular}  &  OT & \OK  &    \\ \cmidrule[1pt]{2-5}
& \begin{tabular}[l]{@{}l@{}} OSSTMM~\cite{Herzog20120OSSTMM}\end{tabular}        & IT & \OK  &    \\ \cmidrule[1pt]{2-5}
& \begin{tabular}[l]{@{}l@{}} CYRail~\cite{cyrail2017SafetySecurityRail}\end{tabular}        & OT &\OK   &    \\\cmidrule[1pt]{2-5}
& \begin{tabular}[l]{@{}l@{}} 4SECURail~\cite{4SECURail}\end{tabular}    & OT  & \OK  &   \\ \cmidrule[1pt]{2-5}
& \begin{tabular}[l]{@{}l@{}} IEC 61508~\cite{iec2010Standard61508}\end{tabular}    & OT  &   & \OK   \\ \cmidrule[1pt]{2-5}
& \begin{tabular}[l]{@{}l@{}} EN 50126~\cite{cenelec2018en50126}\end{tabular}    & OT  &   & \OK   \\ \cmidrule[1pt]{2-5}
& \begin{tabular}[l]{@{}l@{}} EN 50159~\cite{cenelec-en-50159}\end{tabular}    & OT  &  & \OK   \\ \cmidrule[1pt]{2-5}
\end{tabular}
\end{table}

\label{sec:signaling}

\begin{figure*}[!t]
	\centering
	\includegraphics[width=0.92\textwidth]{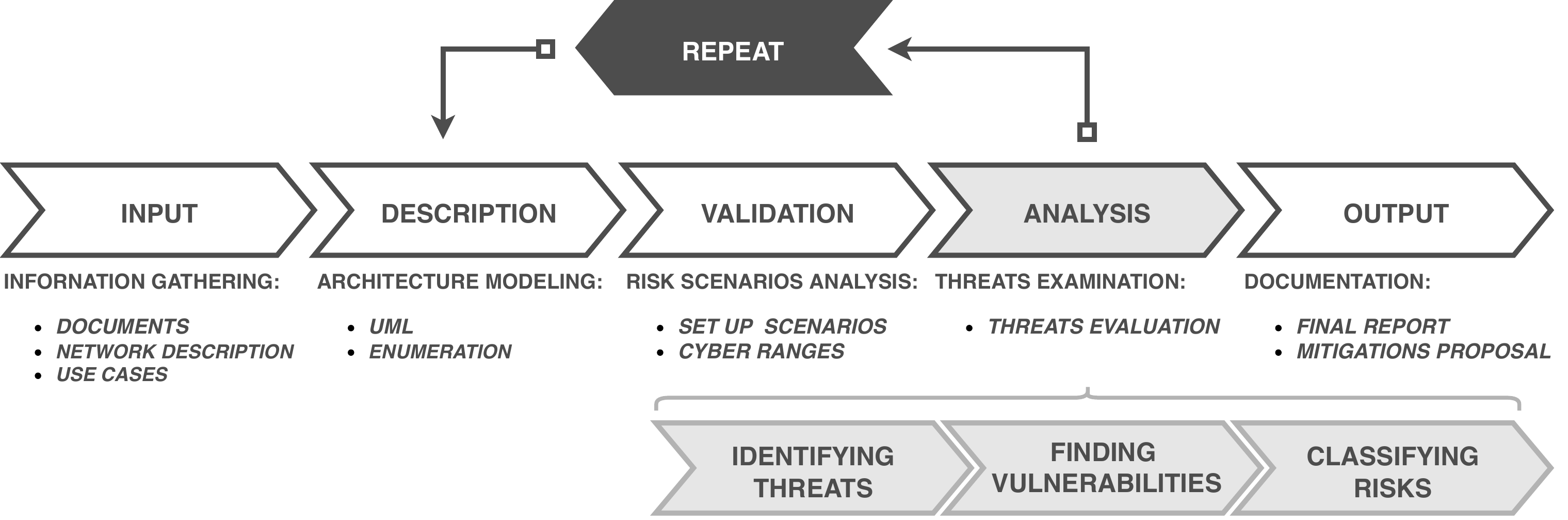}
	\caption{Graphical depiction of the general cybersecurity assessment process detailed in Section~\ref{sec:novelMethodology}. The core stage is highlighted in gray.}
	\label{fig:threats-modeling}
\end{figure*}

\section{An enterprise-oriented approach to security risk management}\label{sec:risk-mgmt}
We have discussed how designers can make their design more secure, but what can a railway operator do to make its \textit{systems as a whole} more secure? In other words, how can operators manage the overall cybersecurity risk?

It is obvious that this task involves both a governance and a technical part, in which the former is tasked to decide the high-level targets involving the security posture of the company and to decide the internal organization  and scope of the cybersecurity teams. The kind of organization usually adopted often follows a hierarchical structure and is tasked to convert the high level decision imposed by higher management into a series of requirements regarding both the materiel and the operating procedure of the company. In turn, these requirements will be usually met by applying the relevant standards, such as the ones we discussed in the previous Sections.

From a technical point of view, the first steps of this process involve performing a \textit{security analysis} of the existing systems to identify weaknesses in the system~\cite{cenelec2021ts50701}. This step will encompass identifying risk scenarios, computing the unmitigated risk, and finally mitigating it.  To do so, the cybersecurity assessors rely on standards and their experience to gauge the strength and effectiveness of the company's security posture.

There are different types of these assessments~\cite{2019:CISSP-guide}. 
For instance, when an organization's internal teams perform such an evaluation, we speak about  \textit{cybersecurity assessment}. Its main goal is to understand the sources of threats, threat events, and possible vulnerabilities on different levels. This process will encompass almost all aspects of a company, spanning from security policies to network architecture and each device's intrinsic characteristics. 
When external experts conduct the analysis, instead, the focus is to measure the compliance of an organization's systems and processes against specific cybersecurity standards and criteria. In such a case, we call this analysis a \emph{security audit}.

In both cases, Cybersecurity Risk Assessments (CRA) will be produced. CRAs categorize cyber risks by likelihood and impact and will be included in a final report directed to the company's management. Such a report will  also be used as a base to write recommendations to improve the security posture of the company~\cite{2019:CISSP-guide}.

It is important to note that, regardless of the actor who performs the investigations, these kinds of processes tend to be highly disruptive to the normal workflow of a company. Consider, for instance, the process of assessing the vulnerability of a given subsystem. This feat will involve automated scans that create considerable traffic load and abnormal interactions in the target systems and probably trigger any system security management tool already present. %
This reasoning is even more actual for penetration tests: this technical methodology extends vulnerability assessments with sanctioned attempts to exploit the discovered vulnerabilities to show their potential real-world impact. In other words, a successful penetration test on live equipment can cause all the negative effects of a real successful attack. 

It is then easy to see that security assessments of any kind  must be seen as a project themselves: clear goals and scope must be established, and operative constraints must be taken into account. A clear communication and cooperation strategy between all parties must also be established to ensure that the overall processes cause no more disservices than absolutely necessary.

\subsection{An applicative procedure for cybersecurity assessments}
\label{sec:novelMethodology}
In this Section, we detail how a security analysis can be carried out. To do so, we take as a test case the network security analysis of a wayside systems\footnote{The scheme we present can also extend to onboard networks.} like the one shown in Figure~\ref{fig:wayside-scenario}.
This procedure can be seen as a summarization of the rules in the standards/guidelines mentioned in Section~\ref{subsec:cybersec}.

The first step of the procedure is the so-called \emph{information gathering} phase. At this stage, one collects information regarding the system under concern, such as requirements, technological assumptions, network characteristics, etc. These data will be used to extract a list of all network components and all interfaces that allow communication between them. Such analysis is the main object of the \emph{architecture modeling} phase.

With such a scheme, one can further proceed with the \emph{risk scenarios analysis}. Depending on the specific case, this step may involve auditing network device configurations, inspecting the policies already in place and real traffic, identifying protocol weaknesses, etc. The analysis of security requirements also takes place at this stage.

The subsequent step is the \emph{threat examination} phase. At this stage, the auditor tries to identify the \textit{threats} that might affect the network under test, namely all those circumstances that might disclose, manipulate, or destroy information, together with all those events that might result in a loss of network availability. This stage, in turn, comprises performing three steps, 
namely:
\begin{itemize}
\item \emph{identifying threats} in software, protocols, and architecture is preliminary for determining the associated risks in the last step;
\item \emph{finding vulnerabilities} in software, protocols, and architecture. Specialized literature, such as the papers mentioned in the previous Section, and public repositories of Common Vulnerabilities and Exposures (CVE)~\cite{cve} list, such as the one overseen by MITRE Corporation, are the most important sources at this stage.
\item \emph{Identifying the associated risk} that derives from the threat event's likelihood and the impact it might have on the network;
\end{itemize}
The cybersecurity assessment ends with a \emph{report}.%

An overall scheme of this procedure is shown in Figure~\ref{fig:threats-modeling}. In the upper part, the main phases of the assessment are shown with a brief explanation of the involved step. In the bottom part, we instead detail the three subphases that comprise the Analysis step.

It is worth mentioning that the threat examination stage resembles the hazard analysis involving hazard identification and related risk analysis and evaluation in the safety assessment process (see~\cite{2018:moving-block-terBeek} and the references therein). However, as described in Section \ref{subsec:safety-security}, the safety and security analyses differ in their focus.

It is also important to remark that, despite being based on the current literature, the procedure we presented in this Section is \textit{novel} in the sense that it does not draw any specific elements from any other existing standard or scientific literature except for its general reasoning and principles. In other words, this is but one among the many possible choices to accomplish a cybersecurity assessment.

\section{Cyber ranges as assessment tools}
\label{sec:ranges}

We mentioned how studying security threats is highly disruptive to do on live systems and possibly very challenging to carry out in a laboratory due to the sheer nature of the required equipment. This is especially true for network-centered subsystems.

With the term \textit{Cyber ranges}~\cite{2018:next-gen-cyberrange}, we indicate all those interactive platforms that allow one to create possibly entirely virtual representations (called  \textit{scenarios}) of existing ICT infrastructures and to emulate their operations by exploiting virtualization and digital twin technologies. Compared to a classical pure experimental laboratory setup, the heavy use of such technology cuts setup and running costs and allows one to operate scenarios also in a totally remote manner.

Scenarios represent a particular combination of active elements, configurations, selected interconnections, and any other specific information required to fully emulate a system. 
It is obvious that to faithfully reproduce complex systems, just like in the case of physical reproductions, a critical challenge is obtaining highly detailed knowledge from the system owners about their systems. This means that, almost ironically, the first benefit one gets when building such virtual scenarios is thus not technological since it forces both owner/operator and security assessors to detail the internal functioning of the original system~\cite{2012:cyber-range-security-assess}.

\subsection{The landscape of cyber ranges}

Many solutions have been proposed to create cyber ranges, depending on the complexity, typology, and purpose of use. To better assess the technological landscape, an idea is to distinguish between cyber ranges based on \textit{simulation-based} architectures from those based on \textit{emulation-based} architectures. The difference is that a simulation environment mimes the essential characteristics of the physical system but neglects low-level implementation details. Such details, however, may be crucial for a thorough network security analysis. Instead, an emulation environment reproduces most physical system peculiarities on a virtual platform. 

When all the scenario components in a theater adopt virtualization solutions to emulate physical devices, some authors classify them as \textit{virtual}. \textit{Physical} theaters provide a replica of the target infrastructure in an isolated and secure environment. \textit{Hybrid} cyber ranges adopt solutions relying on a combination of hardware, virtualized, and simulated elements.

The development and execution of experiments in cyber ranges involve labor-intensive and error-prone operations. For this reason, most cyber ranges platform heavily rely upon automation~\cite{gustafsson2020cyber,2020:CRACK}. 
However, many commonly used off-the-shelf theater automation (also called \textit{orchestration}) frameworks, and, in turn, the cyber ranges that exploit them, do not provide a complete set of network-related functionalities. 
Luckily, various examples of network-centered cyber ranges also exist. 
They, however, widely differ in performance and implementation methods. In Table~\ref{TAB:emulators}, we present a brief schematic comparison between some well-known network emulators\footnote{A complete systematic review of cyber ranges landscape would be outside this work's scope.}.

Among the essential features that one would like to have in a cyber range, the capability of interacting with components that live outside the emulated scenario is of paramount importance. Although this might sound counter-intuitive since the isolation guarantees are one of the biggest selling points of cyber ranges, this feature is also crucial to test the interactions with non-emulable components (such as industrial supervisory control systems). This allows specialists to assess the effect of various otherwise non-easily testable vulnerabilities. We refer the interested reader to~\cite{costa2020automating,2015:SDN-survey} for further reading on the topic and to the PAIDEUSIS~\cite{2021:paideusis} project for an example of successful integration between a virtual environment and physical machinery.

\begin{figure}[!t]
	\centering
	\includegraphics[width=0.99\columnwidth]{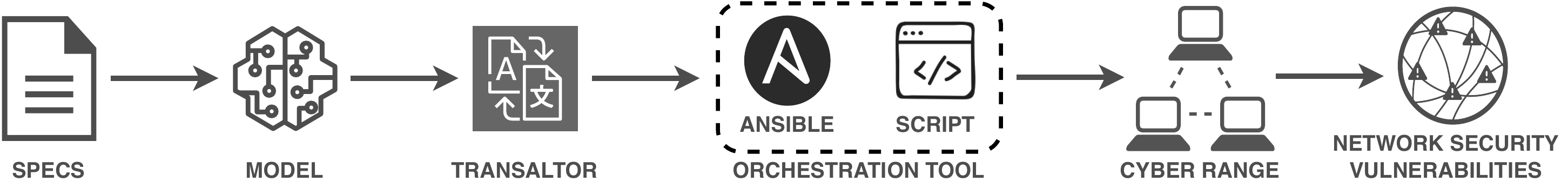}
	\caption{Graphical depiction of the working pipeline adopted to perform vulnerability assessments using a cyber range. }
	\label{fig:cyber-range_data-flow}
\end{figure}

\subsection{The role of cyber ranges for network security}

A use-case in which cyber ranges shine is enabling security testing of complex systems. For interconnected systems, for instance, they allow one to study how malware propagates on its network or to emulate the effect of an attack at L2 on a given link. 

Cyber ranges naturally fit in the workflow presented in Figure~\ref{fig:threats-modeling}. Following the proposed procedure, the ``description phase'' entails enumerating the system components and modeling them in a  cyber range-friendly way,  i.e., in a way that configuration management and orchestration tools can immediately process  (e.g., an specifically crafted YAML configuration file in the case of a cyber range orchestrated using Ansible~\cite{ansible}). If done right, this allows one to create scenarios and straightforwardly analyze the system vulnerabilities.

Security threats can have many faces: flawed network architecture, erroneous device configurations, weak protocols, etc. 
A significant advantage of cyber ranges is that, thanks to their high fidelity simulation capabilities enabled by virtualization technology, they allow practitioners to study how threats combine their effects in a way that would be practically infeasible for other approaches.

Conforming to the cyber kill chain approach (please refer to Appendix~\ref{sec:killchain} for details), the workflow for using a cyber range to evaluate network security broadly considers the following steps:
\begin{enumerate}
    \item emulate the network (or a part of it) using the actual configurations and operating systems of the involved devices;
    \item research the vulnerabilities through automatic tools and scripts created for the specific case;
    \item enumerate the vulnerabilities and measure their impact on the system under test.
\end{enumerate}

Once one has found a vulnerability using the cyber range, the security analyst can proceed in developing a countermeasure and give evidence of it using the very same tools used for assessing the presence of vulnerabilities in the first place. This process will result in new scenarios that no longer contain the vulnerability and can be used for further analysis. %
The procedure can be repeated until the analysis has covered all the network segments.

Figure~\ref{fig:cyber-range_data-flow} depicts the methodology presented in this Section in a flow chart. In there, we stressed the often overlooked importance that orchestration tools assume in making ranges an actual practical instrument.

\begin{savenotes}
	\begin{table*}[t]
		\centering
			
	\caption{ A schematic comparison between some well known cyber-ranges solutions: Cisco Modeling Labs (CML)~\cite{cisco-ml}, Common Open Research Emulator (CORE)~\cite{CORE}, Emulated Virtual Environment - Next Generation (EVE-NG)~\cite{eve-ng}, Graphical Network Simulator 3 (GNS3)~\cite{gns3}, and  Mininet~\cite{mininet}. All these emulators provide means for connecting external nodes, but only CML, EVE-NG, and GNS3 support device operating system virtualization. }
		\label{TAB:emulators}
			\begin{tabular}{@{}clccccc}
				& 
				& \textbf{CORE}
				& \textbf{Mininet}
				& \textbf{EVE-NG}
				& \textbf{GNS3}
				& \textbf{CML}\\
				\cmidrule{2-7}
				& \begin{tabular}[c]{@{}l@{}} \textbf{Network configuration}\end{tabular} 
				& Python, Labs
				& Python, CLI 
				& API, Labs 
				& API, Labs 
				& API, Labs \\ \cmidrule{2-7}
				& \begin{tabular}[c]{@{}l@{}}\textbf{Network emulation level}\end{tabular}
				& \begin{tabular}[c]{@{}c@{}} L3,\\ (L1/L2 EMANE)\end{tabular}
				& L2
				& L2 
				& L2
				& L2 \\ \cmidrule{2-7}
				& \begin{tabular}[c]{@{}l@{}}\textbf{Connection to external nodes}\end{tabular}
				& Yes
				& Yes 
				& Yes 
				& Yes 
				& Yes\\ \cmidrule{2-7}
				& \begin{tabular}[c]{@{}l@{}}\textbf{Nodes operating system emulation}\end{tabular}
				& No
				& No
				& Yes 
				& Yes 
				& Yes \\ \cmidrule{2-7}
				& \begin{tabular}[c]{@{}l@{}}\textbf{Licensing}\end{tabular}
				& BSD
				& BSD 
				& GPL, Commercial
				& GPL 
				& Commercial \\
				\cmidrule{2-7}
			\end{tabular}
		\end{table*}
\end{savenotes}

\subsection{Building cyber ranges: an applicative example for railwyas and signaling} 
\label{sec:cyberangeExample}

\begin{figure*}[!t]
	\centering
	\includegraphics[width=0.99\textwidth]{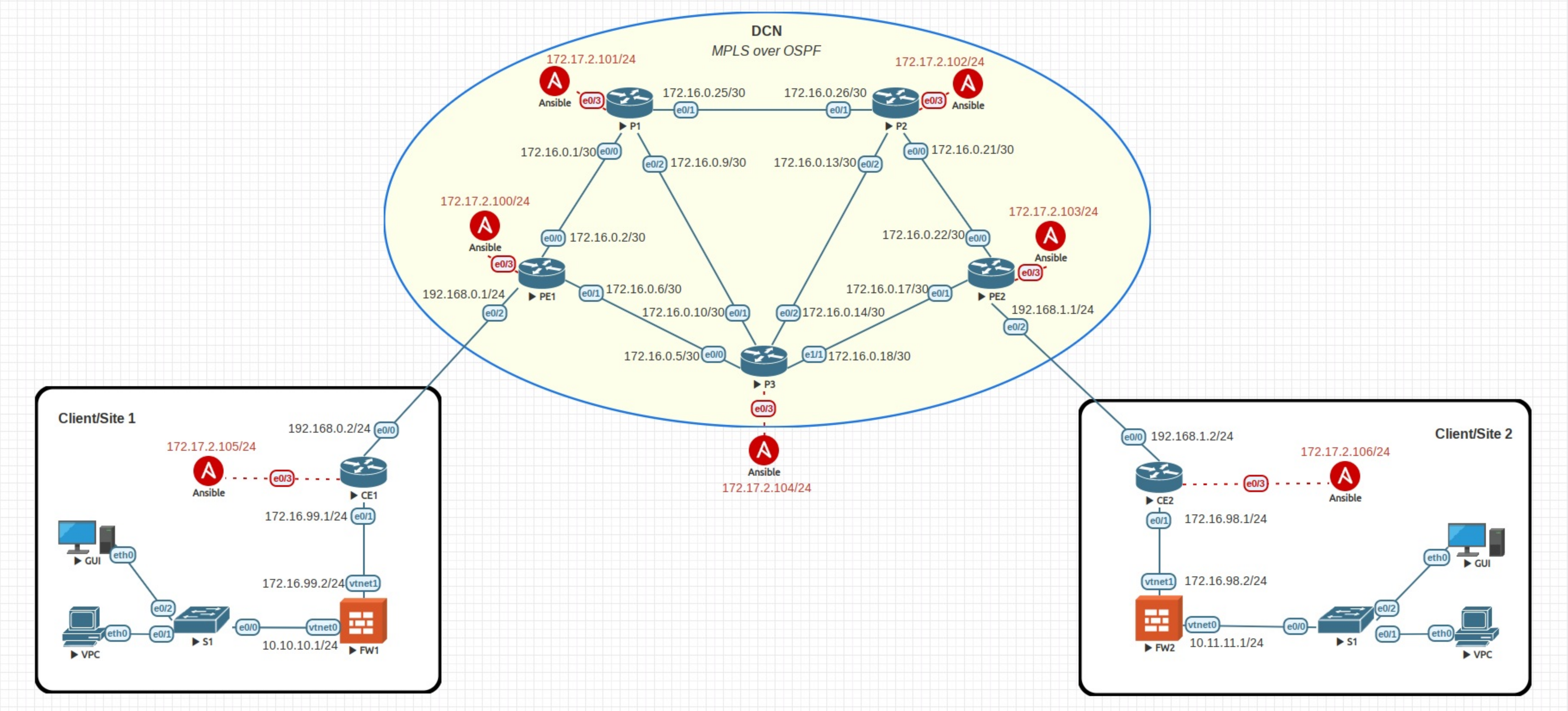}
	\caption{Example of using cyber-range to used to emulate an IP/MPLS network. The picture represents the network described in Section~\ref{sec:cyberangeExample}.}
	\label{fig:cyber-range_mpls}
\end{figure*}

As a practical example of the proposed procedure, in this Section, we show how EVE-NG can be used to investigate an imaginary IP/MPLS\footnote{A proper treatment of MPLS networks is outside the scope of this document. We refer the interested reader to~\cite{DeGhein2016MPLS} for further readings on the topic.} backbone like the one shown in Figure~\ref{fig:cyber-range_mpls}.  

This kind of network is a realistic representation of what a railway operator may use to interconnect equipment in different stations and is inspired by~\cite{CISCO2007MPLSDCN, CISCO2020MPLSVPN}.
In the picture, we can distinguish a central core network representing the core routers of the backbone, possibly connected by high-speed fiber optic links that may span a country. The core is tasked to offer connectivity to, among others, signaling boxes along the rail lines. Connections related to different services (signaling, alarms, etc.) are segregated using MPLS VPN technology, which ensures separation between the traffic generated by the different local area networks inside the signaling boxes. 
In each local network, we can recognize a customer edge router connected to the core and a firewall. This latter device guards the traffic flowing into each local area. 
This scheme describes, for instance, a situation in which many signaling boxes (the clients) are connected to a central OCC (not shown but conceptually identical) using VPN technology using a railway holder own infrastructure.

In the scenario, configurations for each device are managed via Ansible, meaning that configurations can be easily modified and applied to stress different aspects of the network, facilitating the discovery and assessment of vulnerabilities.

We remark that in this shown scenario, the devices are virtualized. Although this may cause more difficulties for the first setup, it also means that each component will actually behave like the real one instead of being a mere reproduction whose functioning may differ due to slightly different implementation details.

Reproducing this scenario in a cyber-range allows one to test the following cases:
\begin{enumerate}[label=Scenario \arabic*, leftmargin=*]

    \item What would happen if an attacker could make one of the links in the core unavailable? Would the internal routing protocol of the core converge again fast enough to guarantee the continuous operation of the overlying systems? \label{item:linkDownConvergence}
    
     \item Would the same conclusion also hold in the case of a \textit{flapping link}? Would the core be able to react fast enough, or would its transient behavior hinder the overlying applications?\label{item:linkDownFlap}%
     
    \item How does a given OT protocol works in case of a congested network? Does the degradation of the links cause violations of timing constraints, loss of information, etc.? 
   
    \item How many resources can be consumed by DoS attacks on IT applications running on a separate VPN but sharing the same infrastructure? Would the necessary separation be maintained, or should QoS policies be implemented to ensure the functioning of OT applications?\label{item:DoS}%
    
    \item Would manipulating a given link by some means allow one to establish a side-channel information transfer? Would this possibly compromise the separation between VPNs? \label{item:VPNSeparation}
    
    \item What would happen if an infrastructure key device is replaced by one from a corrupted node, as in~\cite{chang2016key}? Are other nodes governed by a given protocol capable of circumventing the problem? \label{item:resilience}

    \item Does the overall employed control structure suffer from Zeno behavior~\cite{dolk2016output} following some kind of momentous Denial-of-Service? \label{item:zeno} %

    \item Is the network configuration subject to a given CVE, or is the architecture able to prevent the effect of the exploit?\label{item:exploit}

    \item What kind of application-level performance degradation would a given kind of electromagnetic interference on a device/link cause? \label{item:emi}

\end{enumerate}

Obviously, not every cyber range is the right tool to test \textit{all} those attacks. To test~\ref{item:emi}, for instance, one would require either a strict integration with FEM tools or the ability for the cyber range to work in a hardware-in-the-loop configuration. Similar considerations also arise from scenarios linked to emulating the physical layers, which would be required to practically evaluate solutions like the one in~\cite{gao2021intrusion}: very few tools can do so. This should not sound discouraging: most cyber range suites are remarkably flexible and allow a great degree of customization of the scenario details. To give an example, a tool like EVE-NG can easily handle all the first eight scenarios with little difficulty.

A related thought must be brought up concerning \textit{performance} considerations. It would be foolish, for instance, to expect a simulated environment to match the actual throughput of a couple of enterprise-grade routers connected via an actual fiber optic link. Nevertheless, this does not mean that performance-related investigation cannot be performed using this kind of tool.
Indeed, even leaving aside the practical possibility of empirical ``conversion rules'' between simulated and real-world scenarios, there is no reason to believe that \textit{relative changes} would not be realistically represented. This observation implies, for instance, the cyber ranges' ability to quantitatively estimate the effect of \textit{amplification attacks}: if one discovers in a scenario that an attack has a given amplification factor, it should be pretty straightforward to scale the quantities involved to discover the actual resources (such as bandwidth) that an attacker would need to use to cause problems in the physical world.
Similar considerations can be drawn for all those scenarios (such as~\ref{item:linkDownConvergence} and ~\ref{item:linkDownFlap}) where protocol delays or other algorithmic considerations dominate the behavior of the system. In such a case, the hardware mismatch would cause minimal issues.

The same reasoning also holds when it is needed to establish the \textit{feasibility} of some countermeasures. Suppose, for instance, that we would like to test if solutions like public key infrastructures (such as X.509, although arguably ill-suited for OT applications~\cite{thomas2017traks}) or the proposal like the one~\cite{zhu2021joint} are too computationally heavy to be implemented within real-time constraints. Even if the actual run time of the cryptographic primitives may be distorted by modern CPU hardware acceleration, the number of packets exchanged due to the intrinsic functioning of protocols (such as handshakes) will be faithfully represented in the virtualized scenario. This allows, for instance, to establish if such handshakes are too time-consuming by simply considering the transport delay of a real country-spanning link and the number of packets exchanged.

Although all the aforementioned considerations could also be drawn using physical equipment, on a practical level, cyber ranges are the most cost-effective tool to enable companies to deploy internal cybersecurity teams (with both ``blue'' and ``red'' teams roles), thus allowing a proactive approach toward security.
This capability enables organizations to train cybersecurity response teams to respond to attacks~\cite{Vekaria2021CyberR} carried out by different kinds of attackers (is the attack source internal or external to the network? Is it a one-person job or coordinated state-sponsored action?) in a relatively effortless manner~\cite{terBeek2021102381, Kuhl2007cyberattacksimulation}.
Indeed, the cyber ranges have already shown their potential as invaluable tools for training purposes~\cite{yamin2020cyber,2021:paideusis}: events such as the CyberChallenge.IT~\cite{cyberchallengeit}, which have become possible thanks to such tools, are witnesses to this fact.

\section{Future challenges and conclusions}
\label{SEC:Conclusions}

In this survey, we have reviewed the current landscape regarding the cybersecurity of railway systems, with a special focus on signaling systems and noting how they strongly rely on complex communication networks. %

To do so, we first analyzed the origin of the ever-increasing interest in new tools and methodologies for cybersecurity rapid risk assessment for safety-critical infrastructure and reviewed the guidelines that can be applied to rail signaling scenarios. We then proposed a novel cybersecurity assessment procedure and showed how one can heavily exploit cyber ranges as an enabling technology to create virtual scenarios in which each vulnerability can be tested and its impact/risk assessed. As a result, our assessment procedure can help improve the cybersecurity posture of railway systems by understanding and mitigating cyber threats and vulnerabilities.

Many questions remain open. The most pressing one is how to better integrate cyber ranges with digital twins. This would allow for an unprecedented level of fidelity and allow for simultaneously studying the safety and security aspects of the systems under concern. 
To this end, we will also need to address the computational requirements of high-fidelity simulation environments. Indeed, as the final goal would be to emulate a railway system in its entirety, it is easy to see how the required computational resources might be beyond the current state-of-the-art capability.

Another challenge is definitely on the cultural level: how can we train tomorrow's technicians? Gamification has been used in cybersecurity for many years as an analysis tool, but can it be used for proper training purposes? Cyber ranges can be powerful tools also to this end, so it will be interesting to see if and how railway companies will adapt to such tools.

\appendices

\section{An overview of railway market} 
\label{sec:market}

In order to better understand the scenarios that a company may have to face when assessing the security profile of a railway system, it is helpful to introduce the way railway systems are often classified based on the intended task they are meant to achieve.
The first distinction to be made is between passenger and freight rail services. Among the former, we can further distinguish based on the distance traveled and the kind of territory served (e.g., urban, inter-regional, etc.)
Railway networks are often organized around mainline rails that serve as a route between major urban centers and to which branches, yards, sidings, and spurs are connected. Mainline is used to provide both High-Speed Rail\footnote{An HSR service is defined as a service that achieves a speed~\cite{uic2018hsr} of at least 200 km/h, regardless of the distance covered~\cite{garmendia2012high}.} (HSR) services and conventional speed rail services.
\textit{Regional traffic} may or may not share the infrastructure with mainline traffic~\cite{uic2015vision} and provides conventional medium/short-based services.
Finally, an urban/sub-urban segment may share the tracks with ordinary road traffic and is often separated from the mainline rail traffic. Examples of such traffic are metros, tramways, and light rails.

\section{Incidents and their modeling}

\label{appendix:lista}
Many authors have formally proposed approaches to analyze such cyber attacks to understand better the attack patterns used. We cite here attack graphs~\cite{phillips1998graph}, 
trees~\cite{schneier1999attack,mauw2005foundations}, 
vectors~\cite{mulazzani2011dark},  
surfaces~\cite{manadhata2010attack}, 
over and above diamond model~\cite{caltagirone2013diamond}, 
OWASP threat model~\cite{Drake-threat-model}, and 
the so-called ``kill chain'' approach. 
See, e.g., \cite{alMohannadi2016cyber,kour2022review} for an overview of some of these models and 
\cite{pizzi2020cybersecurity,kour2020railway} for applications of attack-fault trees to analyze some cybersecurity-related incidents in the rail industry.
We also cite~\cite{zhang2021physical,duo2022surveyCPS,zacchialun2019jss} for a reviewer on the security of cyber-physical systems from a control-theoretical prospective.

\section{The cyber kill chain approach}
\label{sec:killchain}

The Kill Chain is a classic military concept that can be used to analyze the structure of an attack. More recently, Locked Martin introduced the same concept in the cyber domain context~\cite{higgins2013lockheed} to better model attacks involving network intrusions. It involves seven phases, each of which has a preferred mitigation approach.

According to the kill chain approach to modeling threats~\cite{assante2015industrial}, cyber \emph{reconnaissance} is the first step an attacker performs when trying to breach a system. There are two types of reconnaissance: passive and active. Passive reconnaissance is when the attacker gathers information about a target without direct interaction. Active reconnaissance is when an attacker directly interfaces with a target system to gather specific details that are later helpful in delivering a malicious payload.
The subsequent phases are~\cite{Albach2019IoT}: \textit{weaponization}, in which the intruder creates remote access malware ``weapons'' tailored to one or more vulnerabilities;
\textit{delivery} of the weapon to the target; \textit{exploitation}, which happens when the ``weapon'' is triggered; \textit{installation}, which refers to the phase in which the weapon installs backdoors of various kind; \textit{command and control}, in which the malware enables an intruder to have access to the target network.
The last phase is referred to as \textit{actions on objective}. Here the attackers take action to achieve their goals.

We refer the reader to~\cite{kour2020railway,assante2015industrial,hutchins2011intelligence} for a more detailed presentation of this topic and how kill chains can be used to analyze cybersecurity-related incidents, also in the rail industry.

\section*{Acknowledgments }
\label{SEC:ACK}
The authors would like to thank Maurice H. ter Beek and Rocco De Nicola for their insightful comments.

\bibliographystyle{ieeetr}
 
\bibliography{biblio}

\end{document}